# Transition of Photonic Dissipative Dynamics through the Exceptional Point

Yue Cui[1], Ruihen Jin[1], Xiong Jiang[1], Kaiyang Wang[1], Can Huang[1,3,5,*], Qinghai Song[1,2,3,4,5,*]

[1] Ministry of Industry and Information Technology Key Lab of Micro-Nano Optoelectronic Information System, Guangdong Provincial Key Laboratory of Semiconductor Optoelectronic Materials and Intelligent Photonic Systems, Harbin Institute of Technology, Shenzhen 518055, China.

[2] Pengcheng Laboratory, Shenzhen 518055, China.

[3] Quantum Science Center of Guangdong-Hongkong Macao Greater Bay Area, Shenzhen 518055, China.

[4] Collaborative Innovation Center of Extreme Optics, Shanxi University, Taiyuan 030006, Shanxi, China.

[5] Heilongjiang Provincial Key Laboratory of Advanced Quantum Functional Materials and Sensor devices, Harbin Institute of Technology, Harbin 150001, China.

*Corresponding author: huangcan@hit.edu.cn; qinghai.song@hit.edu.cn

**Abstract**

The decay of light in an optical structure depends not only on the intrinsic properties of the material but also on the surrounding electromagnetic environment. This principle has laid the foundation for the engineering of dissipation in photonic emitters. In the conventional wisdom, dissipation is governed by a fixed set of decay channels, each defined by the eigenstates of the structure, and energy leaks through and interacts with these channels. Here we provide experimental evidence that this paradigm fails in non-Hermitian systems. Specifically, we observe an accelerated transient decay in a pair of coupled microcavities tuned near the exceptional point, revealing that photonic dissipation can be governed not by reconfiguring existing loss channels, but rather by restructuring the underlying state space. The universality of this phenomenon is corroborated through two independent control parameters.The finding provides a new perspective on dissipative dynamics in open optical systems and offers a distinct mechanism for controlling transient decay in ultrafast photonic systems.

## Introduction

Controlling how light escapes a structure is one of the oldest problems in optics. Since Purcell recognized that spontaneous emission is not an immutable atomic constant but a response to the electromagnetic environment [1–3], the engineering of dissipation has been synonymous with the engineering of spectra: reshaping the local density of optical states, displacing resonances, raising or lowering quality factors [4-7]. For example, dissipation can be suppressed by placing an emitter inside a photonic-bandgap material, enhanced by coupling it to a high-Q, small-mode-volume cavity, or continuously shifted by tuning the cavity length or refractive index. However, these schemes share a silent premise: The set of decay channels is taken as fixed and complete; engineering redistributes how rapidly each channel leaks, but it can neither change their number nor make one of them vanish. Whether dissipation could instead be controlled by restructuring the state space itself has rarely been explored.

Open systems violate this premise on their own terms. Any resonator that radiates or absorbs is open, and its resonant frequency becomes a complex number, with the imaginary part encoding the decay rate. Especially the exceptional points (EP), which represents non-adiabatic degeneracies in the parameter space of non-Hermitian systems, characterized by the simultaneous coalescence of both eigenvalues and eigenvectors [8–13]. At an EP, the system is no longer described by independent exponential decays but by a single Jordan-block evolution [14,15]. Dissipation is thereafter governed by the coalescence geometry. These unique attributes have unveiled tremendous applicative potential across multiple domains, including enhanced sensing sensitivity, asymmetric mode coupling, and unidirectional wave transport et.al [16-18]. Nevertheless, despite significant advances in EP physics, the vast majority of experimental investigations have remained confined to the steady-state regime, the definitive dynamical signature has not been quantitatively verified in the time domain.

Here we report EP-controlled dissipative dynamics in a femtosecond-pumped coupled-cavity lasers, in which the resonators are defined by the pumping defined quasi-BIC mode. The coupling strength and gain imbalance are independently tuned via the pump-spot separation and relative pulse delay, providing in-situ continuous control over the parameter space for systematic EP crossing [19–21]. Time-resolved measurements reveal that the lasing decay rate follows a scaling law analogous to the steady-state EP signature, while the slow spontaneous-emission background remains invariant. This directly identifies the acceleration as a manifestation of eigenvectors collapse rather than any material property of the gain medium, opening a new route for the active manipulation of nanophotonic emission dynamics in the time domain.

## Theoretical analysis

We start from a general theoretical analysis based on the coupled-mode equation. Consider two coupled resonators, the dynamics of the system can be described by[22]:

$$\frac{dE_1}{dt} = -i\omega_0 E_1 + (\gamma_1 - k_1)E_1 - iJE_2 \tag{1}$$

$$\frac{dE_2}{dt} = -i\omega_0 E_2 + (\gamma_2 - k_2)E_2 - iJE_1 \tag{2}$$

where $E_{1,2}$ denote the complex valued electric field amplitude, $\gamma_{1,2}$ and $k_{1,2}$ represent the gain provided by the pump and the cavity dissipation, respectively. J describes the conservative coupling between cavities. Assume $k_1=k_2=k$, The eigenvalues of the system are：

$$\lambda_{\pm}=-i\omega_0+(\bar{\gamma}\text{-k})\mp i\sqrt{J^2-\Delta^2} \quad (3)$$

where the average gain is: $\bar{\gamma}=\frac{\gamma_1+\gamma_2}{2}$ and the gain difference is: $\Delta=\frac{\gamma_1-\gamma_2}{2}$. At the exceptional point, $\sqrt{J^2\text{-}\Delta^2}=0$, the eigenvalues and eigenvectors coalesce, the Hamiltonian reduces to a Jordan block. Away from the EP, the system exhibits distinct dynamics depending on the sign of $J^2\text{-}\Delta^2$, In the PT-symmetric regime ($J^2\text{-}\Delta^2>0$), the eigenmodes have different oscillation frequencies but identical decay rates. In the PT-broken regime ( $J^2\text{-}\Delta^2<0$), the eigenmodes share the same frequency but decay with different rates. In the latter case, the intensity evolution follows a bi-exponential form [23]:

$$I(t)\propto A_1e^{2\mathrm{Re}(\lambda_+)t}+A_2e^{2\mathrm{Re}(\lambda_-)t} \quad (4)$$

where the two distinct rates originate from the non-orthogonal eigenmodes. At the EP, the two eigenvectors coalesce and the system reduces to a Jordan block. The intensity then exhibits a modified decay：

$$I(t)\propto(1+\alpha t+\beta t^2)e^{-2\Gamma_{EP}t} \quad (5)$$

Where $\Gamma_{EP}$ denotes the decay rate at the exceptional point, and the coefficients α and β are real parameters determined by the initial excitation condition of the two coupled modes. The framework outlined above is well established in theory, yet its experimental verification has been hindered. In passive platforms, the coupling strength is fixed once the device is fabricated, and the loss contrast is dictated by material absorption, which provides limited and indirect control over the system's proximity to the EP [24]. Moreover, the intrinsically fast optical decay in passive cavities precludes the capture of transient dynamics prior to the EP signatures.

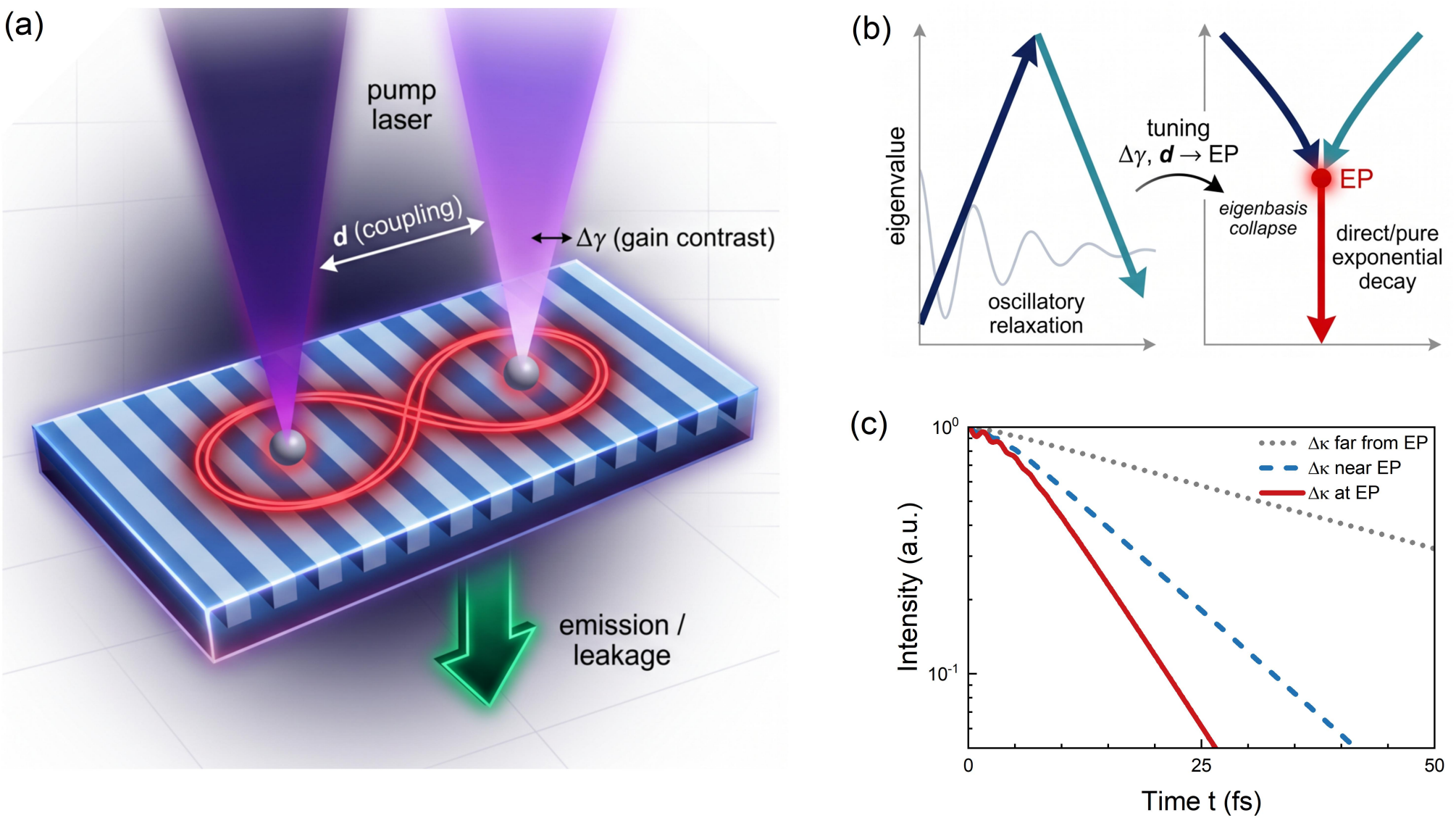

**Figure 1. Dissipative decay channels and dynamical analysis of the linear EP system.** (a) Schematic diagram of experimental setup for two coupled active q-BIC resonators. (b) Schematic diagram of the transient relaxation acceleration induced by the coalescence of dissipative channels at the exceptional point (EP). (c) Comparison of dynamical evolution in different states.

Here, we overcome these barriers by utilizing quasi-BIC lasing resonators, where the coupling strength and the gain imbalance become independently and continuously tunable in situ, thereby enabling systematic traversal across the EP. Crucially, introducing an active gain medium is not merely a means to compensate for cavity loss. The sustained stimulated emission continuously replenishes the coherent field, allowing the transient dynamics near the EP to be resolved on experimentally accessible timescales. To map this active laser system onto the non-Hermitian EP framework, we adopt the coupled Maxwell–Bloch equations as the starting point [25-28]:

$$\frac{dE(t)}{dt}=-k\cdot[E(t)+A\cdot P(t)] \tag{6a}$$

$$\frac{dP(t)}{dt}=-P(t)-E(t)\cdot D(t) \tag{6b}$$

$$\frac{dD(t)}{dt}=\gamma[1-D(t)+E(t)\cdot P(t)] \tag{6c}$$

where E, P, and D are the normalized field amplitude, polarization, and population inversion, respectively, and $\kappa$ and $\gamma$ are the normalized loss and carrier relaxation rates. Near the lasing threshold, the polarization and population inversion relax much faster than the optical field. Adiabatically eliminating these fast variables via the center-manifold reduction [27] yields a Stuart-Landau oscillator description for each cavity. Therefor the nonlinear dynamical equations of the coupled quasi-BIC microlasers read (detailed derivation in SI):

$$\frac{dE_1}{dt}=[(\gamma_1-k)-i\omega-|E_1|^2]E_1-iJE_2 \tag{7a}$$

$$\frac{dE_2}{dt}=[(\gamma_2-k)-i\omega-|E_2|^2]E_2-iJE_1 \tag{7b}$$

Compared with the linear equations (1-2), the nonlinearity here arises from the gain-saturation terms $|En|^2En$, which help to describe the steady-state limit cycle (LC) once the lasing field builds up. Then we can concretely elucidate the dynamical time domain evolution of the coupled lasing system. However, the definition of an EP relies on a linear operator. The presence of nonlinear terms would shift the EP position [28]. Directly using the nonlinear equations to map the time domain phase diagram would obscure the exact dynamical features near the EP. Considering the system operates near the lasing threshold, where the field amplitudes are small and the nonlinear terms are weak. Equation (7) then reduces to its linear form. We therefore adopt a two-step strategy: first we linearize the system at the origin to map the parameter space, revealing the global phase boundaries to find the EP locus, then we use the parameters identified by this linearization to integrate the full nonlinear equations and verify the predicted time-domain dynamics. Specially, we note the coupled quais-BIC laser resonators possess a specific parameter regime where the optical fields decay monotonically to zero—the so-called lasing self-termination or amplitude death (AD) regime [17–19]. Within this regime, the saturation terms $|En|^2En$ become higher-order infinitesimals throughout the entire relaxation process, providing an ideal window for mapping the nonlinear dynamics onto the linear EP theory over the full decay trajectory.

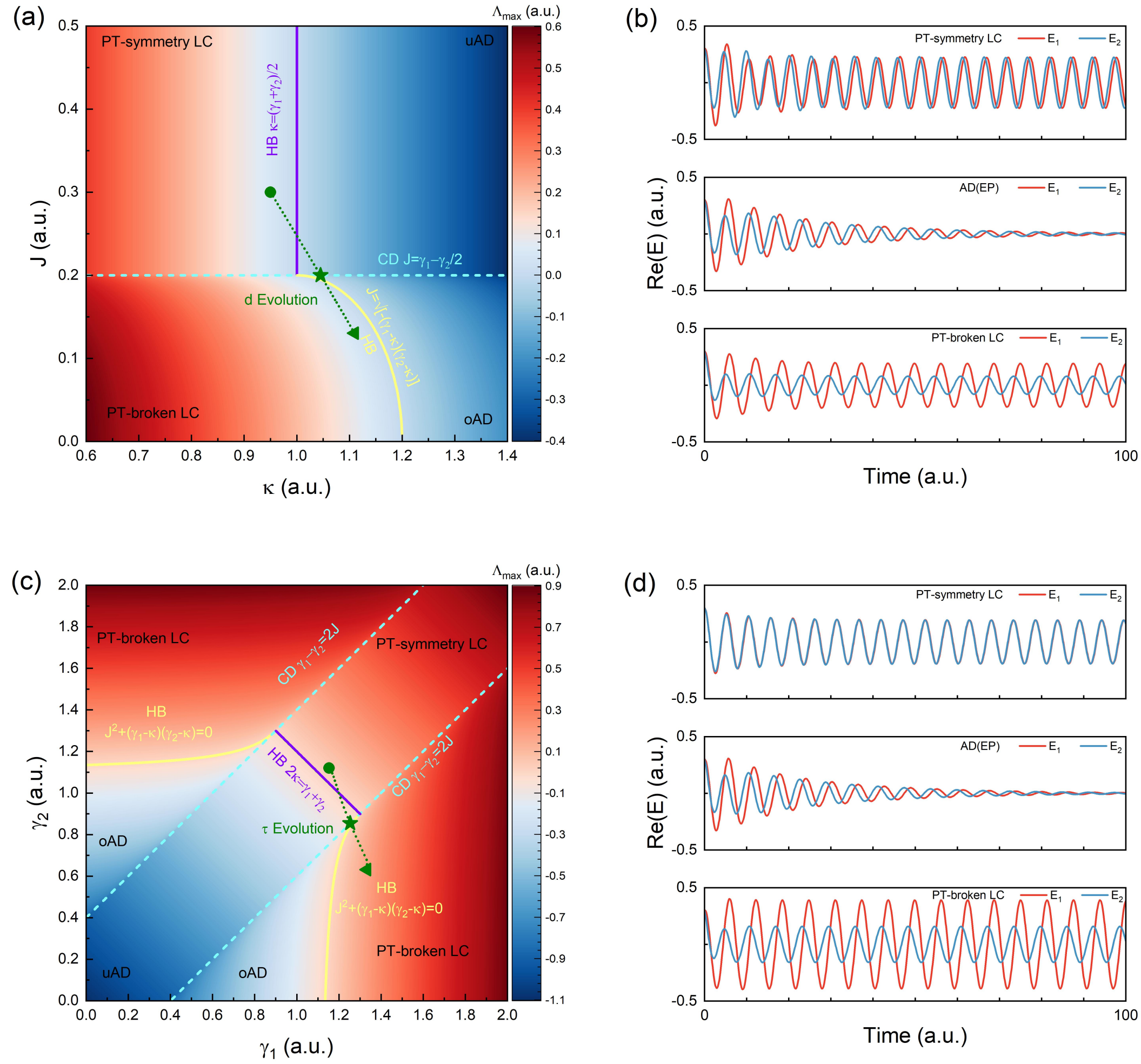

**Figure 2. Global Bifurcation and Dynamical Analysis of Nonlinear EP Systems.** (a) Nonlinear stability theoretical analysis in the κ-J parameter space; parameter settings: $\gamma_1=1.2, \gamma_2=0.8, \omega=1$. (b) Representative examples of dynamical evolution portraits under κ-J parameter tuning (d Evolution); parameter settings: PT-symmetric limit cycle (LC) at (κ=0.95, J=0.3); amplitude death at EP (AD/EP) at (κ=1.05, J=0.2); PT-broken LC at (κ=1.11, J=0.13). (c) Nonlinear stability theoretical analysis in the $\gamma_1$-$\gamma_2$ parameter space; parameter settings: κ=1.1, J=0.2, ω=1. (d) Representative examples of dynamical evolution portraits under $\gamma_1$-$\gamma_2$ parameter tuning: PT-symmetric limit cycle (LC) at ($\gamma_1=1.15$, $\gamma_2=1.12$); amplitude death at EP (AD/EP) at ($\gamma_1=1.25$, $\gamma_2=0.85$); PT-broken LC at ($\gamma_1=1.33$, $\gamma_2=0.63$).

To map the global stability of the system, we linearize the coupled-mode equations (7) around the origin. The stability of this fixed point is determined by the real parts of the eigenvalues $\lambda_\pm$ in Eq. (3). We define the local growth rate $\Lambda_{max} = \max[\mathrm{Re}(\lambda_+), \mathrm{Re}(\lambda_-)]$. When $\Lambda_{max} > 0$, the origin is unstable and the system evolves toward a limit cycle (lasing). When $\Lambda_{max} < 0$, the origin is stable and the system decays to zero (AD). The boundary $\Lambda_{max} = 0$ is defined as the Hopf bifurcation (HB) line, which marks the critical boundary between the AD and LC phases. In Fig. 2a, the HB line is divided into two branches by the EP line: bifurcation in the strong-coupling regime ( $J > |\Delta|$ ) yielding HB1: $\kappa=(\gamma_1+\gamma_2)/2$ . And bifurcation in the weak-coupling regime ( $J < |\Delta|$ ) gives HB2: $J = \sqrt{-(\gamma_1-\kappa)(\gamma_2-\kappa)}$ (detailed derivation see Supplementary section B3).

As shown in figure 2a and 2c, in both κ-J and $\gamma_1$-$\gamma_2$ parameter space, these two bifurcation lines divide the parameter space into four dynamical phases: (i) PT-symmetric LC: in which strong coupling homogenizes the gain difference; the two oscillators lock to equal amplitudes and a fixed phase (see top panel, figure 2b and 2d). (ii) PT-broken LC: in which weak coupling cannot overcome the gain contrast; the supermode localizes on the stronger oscillator, and the weaker one undergoes slave oscillation (see bottom panel, figure 2b and 2d). (iii) Underdamped AD (uAD): in which energy oscillates between the two cavities during decay, producing an inward spiral trajectory in phase space. (iv) Overdamped AD (oAD): in which energy is transferred from the above-threshold oscillator to the below-threshold one and rapidly dissipates, yielding faster decay than uAD. Between the two AD branches lies the critical-damping (CD) line, which is the boundary where the system achieves the fastest decay without the oscillatory bottleneck of underdamping or the slow-mode drag of overdamping. We note the CD line coincides with the EP line, confirming that the coalescence of decay channels at the EP maximizes the decay rate, as shown in the middle panels in figure 2b and 2d.

These phase diagrams serve as a roadmap for the experiment. For example, the two orthogonal tuning paths, marked by the green curves in Figs. 2a and 2c, correspond respectively to varying the coupling strength J and loss factor κ, or the the gain difference Δγ. Along either path, the system traverse dynamical phases, from PT-symmetric lasing, through the AD regime, to PT-broken lasing. Below, we present experimental measurements showing the time-domain dynamics along both paths.

## Experimental verification

The optical setup is illustrated in Fig. 3a. A femtosecond pump beam (Spectra Physics) was split into two paths with independent delay lines and focused onto two selected positions within the quasi-BIC sample. A scanning galvanometer compensated for the relative temporal delay. Inset in figure 3a shown the SEM image of the sample, which consisted of a 100-nm-thick perovskite film deposited on a fused-silica substrate, covered by a polymer layer with a two-dimensional array of nanoapertures with period 340 nm and aperture diameter 110 nm. The pump-spot separation $d$ between the two excited apertures was calibrated by imaging the pump spots onto a CCD camera. The emission was collected and split into two channels: one to a spectrometer (Ocean Insight FX-XSR) for steady-state spectra and the other to a streak camera (XIOPM 5200) for time-resolved transients. A spatial aperture in the collection path selected only the BIC-mode emission (see SI for details).

We first set the pump intensities above (1.2 Pth) and below (0.8 Pth) the single-cavity lasing threshold, with the two cavity centers separated by $d$ = 10 μm. Under this condition, the system resides the PT-symmetric phase and exhibits dual-mode lasing, as shown in Fig. 3b. Keeping both pump powers fixed at these values, we gradually increased the separation $d$. Notably, varying $d$ continuously tunes the coupling strength J between the two quasi-BIC resonators and the intrinsic loss k (see SI for details). As $d$ increases, the steady-state laser spectra (Fig. 3b) evolve from a double-peaked profile (PT-symmetric phase) to a region of strongly suppressed lasing intensity, the amplitude death (AD) regime, and finally to a single-peaked profile (PT-broken phase). The white solid curve in Fig. 3b represents a theoretical fit based on Eq. (7), locating the EP at $d$ = 28 μm within the AD region.

In parallel with the steady-state spectral measurements, a streak camera recorded the time-resolved emission intensity over a 300 ps window for each pump pulse. To extract the intrinsic decay dynamics free from instrumental broadening, we applied a deconvolution

procedure(see SI for details), which separates the fast decay component ($\tau_{lasing}$), associated with the coherent lasing field, from the slow fluorescence background ($\tau_{fluorescence}$). The corresponding dynamic decay rates are subsequently determined by $\Gamma_{lasing} = -1/\tau_{lasing}$ and $\Gamma_{fluorescence} = -1/\tau_{fluorescence}$. Figure 3d shows the normalized decay traces at three representative separations: $d$ = 10 μm (PT-symmetric phase), $d$ = 28 μm (near the EP within the AD region), and $d$ = 60 μm (PT-broken phase). At $d$ = 28 μm, the lasing field decays with a rate $\Gamma_{lasing} = -0.33712$ ps$^{-1}$, markedly faster than at $d$ = 10 μm ($\Gamma_{lasing} = -0.22152$ ps$^{-1}$) and at $d$ = 60 μm ($\Gamma_{lasing} = -0.06283$ ps$^{-1}$), consistent with the accelerated decay expected at the EP. It is worth emphasizing that the absence of steady-state laser emission within the AD region does not imply the absence of a lasing field altogether. Each femtosecond pump pulse generates a transient coherent field, which subsequently decays exponentially to zero over tens of picoseconds. The streak camera, gated to this early time window, captures the complete decay trajectory, whereas the spectrometer integrates over milliseconds and registers only the time-averaged steady-state signal—which vanishes in the AD regime. The AD region is therefore 'dark' only in the steady-state sense, not in the time domain.

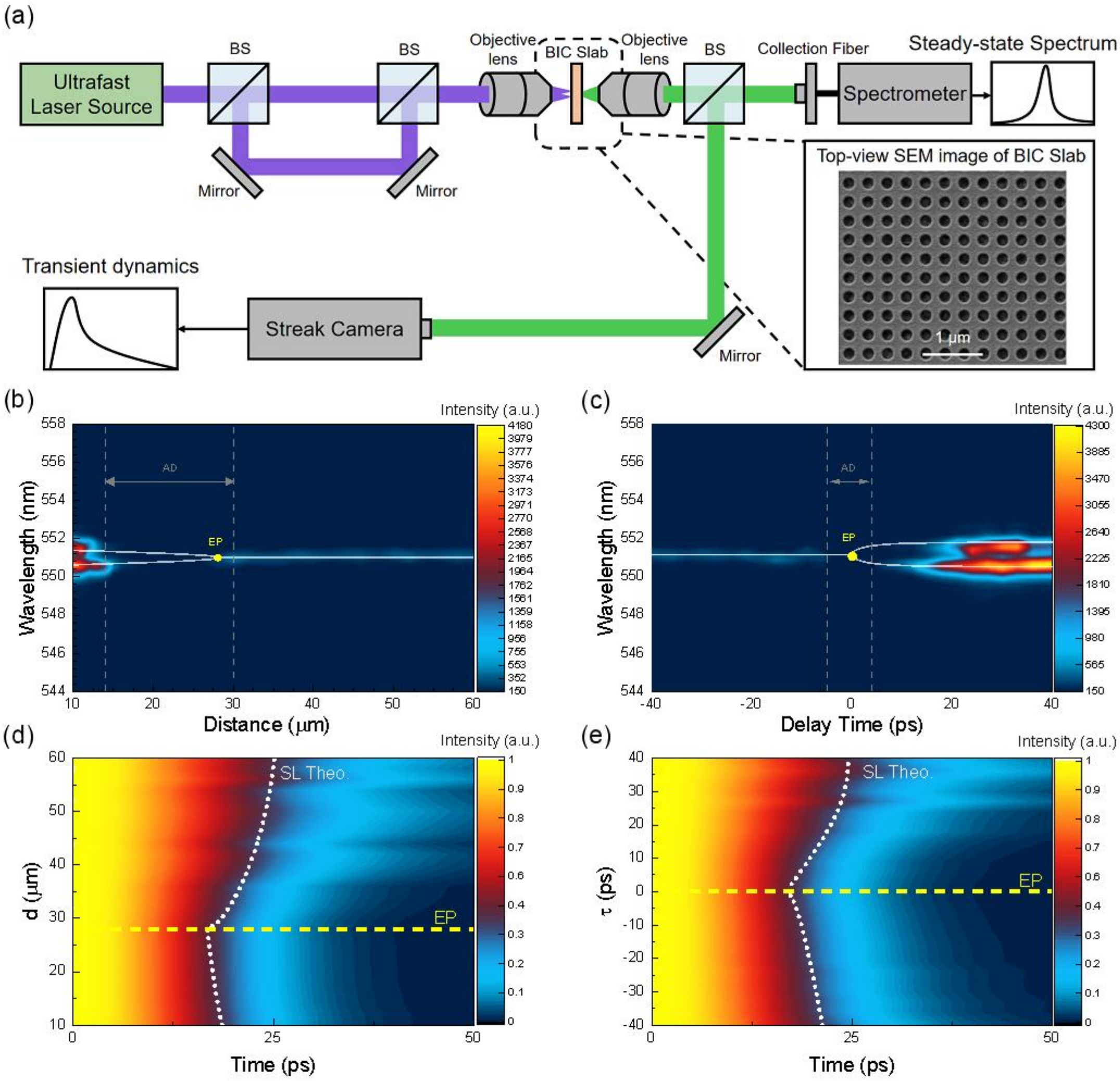

**Figure 3. Verification of Exceptional Point-Accelerated Dissipative Photonic Dynamics.** (a) Optical setup; the inset shows the SEM image of the perovskite-BIC sample. (b-c). Experiment results for stable spectrum measurement when scan the distance and delay time between two pump beams, respectively. (d) Transient

temporal evolution data of the coupled laser system at distances d = 10, 28 and 60 μm, respectively. (e) Transient temporal evolution data of the coupled laser system at delay time $\tau$ = -40, 0 and 40 ps, respectively.

The relative pump delay $\tau$ provides a second, independent tuning pathway, since a cavity pumped earlier has a lower remaining population inversion when the other arrives. The effective gain coefficients $\gamma_1$ and $\gamma_2$ during the mutual interaction therefore depend on the arrival times of the pump pulses at the two cavities. For this pathway, we fixed the cavity separation at d = 28 μm (the EP separation identified above) and the pump intensities at 1.2 Pth and 0.8 Pth, respectively. We then scanned $\tau$ from −40 ps to +40 ps (positive delay means the high-power pulse arrives first, details see SI). When the weak pulse arrives much earlier than the strong one ($\tau \ll 0$), the weakly pumped region has already lost most of its inversion by the time the strong pulse excites the other cavity, it remains below threshold and does not lase, yielding a single-peaked spectrum from the high-power cavity. Conversely, when the strong pulse arrives slightly earlier ($\tau > 0$ ps), photons from the strongly pumped cavity propagate to the weakly pumped one via the BIC waveguide mode, acting as an external seed that raises its effective gain above threshold—allowing both cavities to lase and couple, and producing a double-peaked spectrum characteristic of the PT-symmetric phase. As $\tau$ is scanned from negative to positive, the steady-state spectra evolve from single-peaked through an AD window (around $\tau = 0 \pm 5$ ps) to double-peaked (Fig. 3c), mirroring the phase progression observed along the spacing-tuning pathway. The white curve in Fig. 3c is a theoretical fit from Eq. (7), which locates the EP within the AD region(around $\tau = 0$). Representative time traces at $\tau = -40$ ps (PT-broken), $\tau = 0$ ps (near EP), and $\tau = 40$ ps (PT-symmetric) are shown in Fig. 3e. At $\tau = 0$ ps, the lasing decay rate reaches $\Gamma_{lasing}$=−0.34526 ps$^{-1}$, in close agreement with the value obtained via the spacing-tuning pathway. That two independent and physically distinct tuning mechanisms converge on the same extremal decay rate at the EP provides strong evidence for the reproducibility and generality of the observed phenomenon.

We further analyze the dynamical scaling behavior of the system. Due to the highly degenerate eigenspace at the EP, the system exhibits extreme sensitivity to small perturbations[29,30]. Figure 4a summarizes the relationship between the pump-spot separation *d* and the corresponding decay rate $\Gamma_{lasing}$ (blue circles). On the PT-symmetric side ($d < 28$ μm), the $\Gamma_{lasing}$ becomes progressively more negative approximately linearly with increasing *d*, indicating that the system relaxes progressively faster as the two cavities are brought toward the EP separation. At $d = 28$ μm, $\Gamma_{lasing}$ reaches its most negative value (i.e., the largest decay-rate magnitude), identifying this extremum with the EP located within the AD region in Fig. 3b. Beyond this point, on the PT-broken side ($d > 28$ μm), $\Gamma_{lasing}$ becomes less negative nonlinearly with further increasing *d.*

Superimposed on the experimental data (red solid line) is the theoretically calculated local growth rate $\Lambda_{max} = \max\,[\mathrm{Re}(\lambda_+), \mathrm{Re}(\lambda_-)]$, mapped onto the same *d*-axis through the parameter calibration between *d* and (J, k) (see SI). The agreement between $\Gamma_{lasing}$ and $\Lambda_{max}$ is quantitative across the entire tuning range: both reach their extrema at the same *d*, and their *d*-dependent variations follow the same functional form. This correspondence is not a fitting result, the theoretical curve contains no parameters fitted to the decay data, and therefore provides direct experimental confirmation that the transient dynamics in the AD regime are governed by the same linearized eigenmode structure that defines the EP.

We attribute the above scaling behavior to the non-Hermitian degeneracy of the eigenmode structure at the EP. Away from the EP, the two eigenmodes, though decaying at different rates , are generically non-orthogonal and therefore both get excited by any realistic initial condition. The overall decay is bottlenecked by the slower channel, while the faster component is masked by the signal background. At an EP, by contrast, the coalescence of the two eigenmodes eliminates one degree of freedom altogether, and the system no longer possesses two independent decay channels to average or bottleneck between, but reduces to a single Jordan-block mode governed solely by $\Gamma_{EP}$. This reduction manifests as a faster observable transient decay, providing a direct manifestation of EP-driven dissipative acceleration.

Crucially, the slow fluorescence decay rate $\Gamma_{fluorescence}$ (gray dots in figure 4a and 4c) remains essentially constant at approximately -0.01 ps$^{-1}$ across the entire *d* range. Unlike the lasing emission, which is directional and efficiently couples into the BIC mode, fluorescence is emitted isotropically; only a small fraction is captured by the BIC mode. The measured fluorescence background does not reflect the non-Hermitian coupling dynamics of the BIC channel, and rules out a trivial Purcell-type enhancement of single-cavity loss as the origin of the observed acceleration.

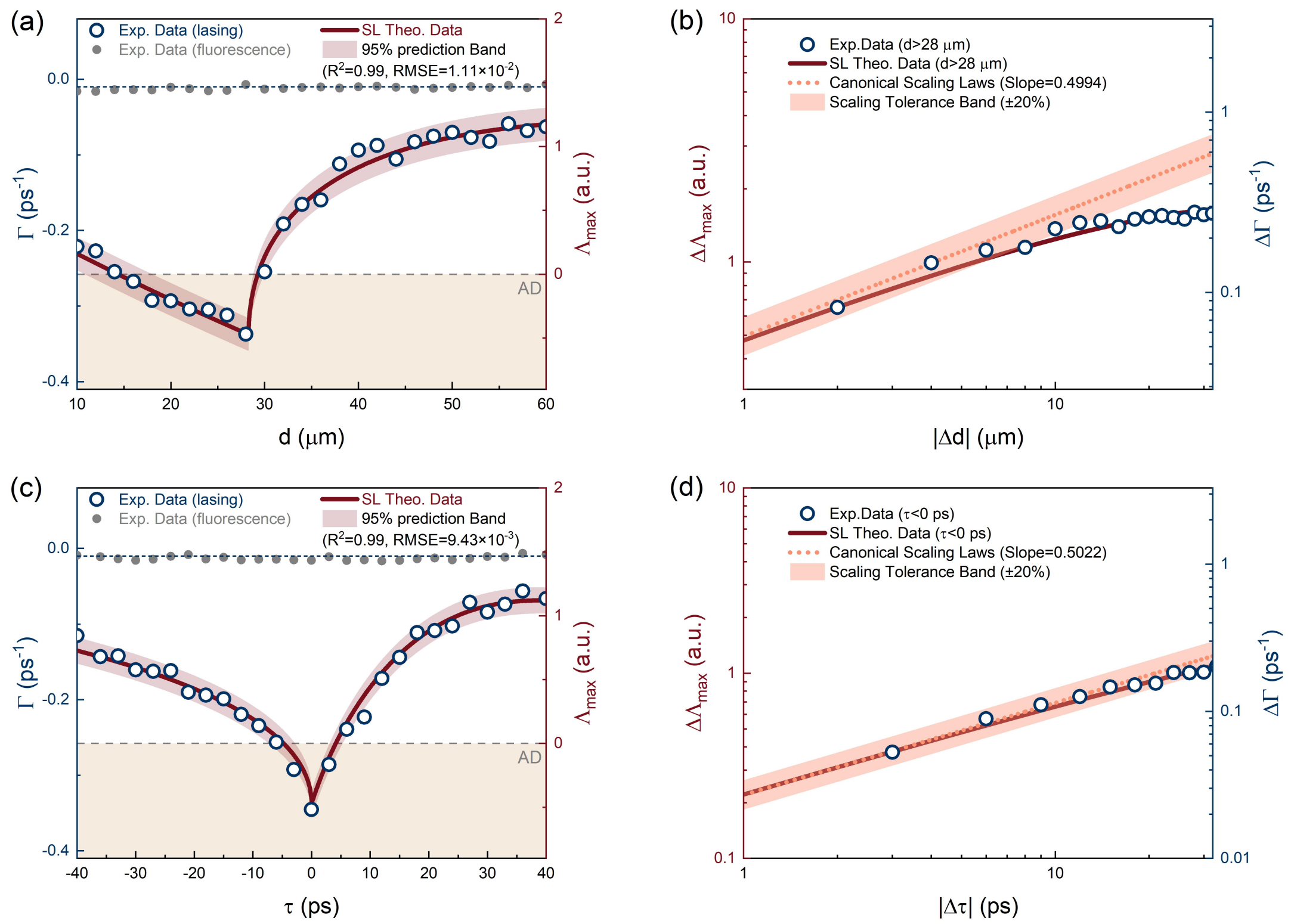

**Figure 4.** $\Lambda_{max}$ **mapping and square-root scaling laws of EP-driven dissipative photodynamics**.(a) Parameter evolution of $\Gamma_{lasing}(d)$ and its linear mapping to $\Lambda_{max}(d)$. (b) Square-root scaling law relationship between $\Delta\Gamma(d)$, $\Delta\Lambda_{max}(d)$ and $\Delta d$ on double-logarithmic coordinates. (c) Parameter evolution of $\Gamma_{lasing}(\tau)$ and its linear mapping to $\Lambda_{max}(\tau)$ (d) Square-root scaling law relationship between $\Delta\Gamma(\tau)$, $\Delta\Lambda_{max}(\tau)$ and $\Delta\tau$ on double-logarithmic coordinates.

Notably, the coalescence of eigenvalues produces a characteristic square-root dependence of the complex eigenfrequency splitting on the perturbation strength near an EP [30,32]. In the time domain, this translates into a square-root dependence of the decay rate deviation on the parameter deviation from the EP. To test this prediction, we define the deviation of the fast decay rate from its value at the EP, $\Delta\Gamma = \Gamma_{lasing}(d) - \Gamma_{EP}$and the corresponding parameter deviation $\Delta d = d - d_{EP}$. In the double-logarithmic plot of Fig. 4b, the SL theoretical data near the EP on the PT-broken side ( $d>d_{EP}$ ) fall on a straight line with a slope of 0.4994, following the predicted scaling $\Delta\Gamma=\Gamma_{lasing}(d)-\Gamma_{EP}\propto\Delta\Lambda_{max}\propto\Delta d^{1/2}$. This square-root dependence is the time-domain analog of the well-known frequency splitting near an EP [30,32], and it directly confirms that the observed acceleration is governed by the Jordan-block geometry of the coalescing state space rather than by any trivial parameter tuning.

Besides, the delay-tuning pathway, which varies the gain difference rather than the coupling strength, provides an independent test of the universality of this scaling behavior. Figures 4c and 4d summarize the results of this second pathway. In Fig. 4c, $\Gamma_{lasing}(\tau)$ (blue circles) again exhibits a minimum at $\tau = 0$ ps and tracks the theoretical $\Lambda_{max}(\tau)$ (red solid line) across the entire scanned range. The fluorescence background $\Gamma_{fluorescence}$ (gray dots) remains unchanged, consistent with the interpretation that the effect is selective to the coherent lasing component. In the double-logarithmic plot of Fig. 4d, $\Delta\Gamma = \Gamma_{lasing}(\tau) - \Gamma_{EP}$ follows the same square-root scaling, $\Delta\Gamma \propto \Delta\tau^{1/2}$, the SL theoretical data near the EP on the PT-broken side ($\tau<\tau_{EP}$) fall on a straight line with a slope of 0.5022. The consistency between the two orthogonal tuning pathways: one in real space (cavity separation, varying J and k) and one in time (pulse delay, varying $\Delta$), confirms that the square-root scaling is not an artifact of a particular control parameter, but a universal signature of the EP-induced restructuring of dissipative channels.

It should be noted that, the SL normal-form model relies on the near-threshold assumption, which permits the adiabatic elimination of the medium polarization P and population inversion D. As the gain difference increases, the system deviates further from threshold, and the fast variables P and D, which were originally slaved by the slow field variable, gradually regain their influence and begin to dominate the field dynamics to some extent. To quantitatively delineate the valid applicability boundary of the near-threshold Stuart-Landau (SL) normal-form model, we further tested the correspondence between the laser decay rate and the perturbation parameters in the vicinity of the non-AD region, specifically, by fixing the total pump energy and increasing the gain difference between the two oscillators. We found the continued increase in gain difference causes the experimentally measured decay rate response curve to shift upward overall and gradually approach a flat plateau. Ultimately, this plateau corresponds to the dynamical response of the first above-threshold laser, essentially marking the applicability boundary of our theoretical framework (see Supplementary Information for details).

## Discussion and Conclusion

The central message of this work is not that an exceptional point can accelerate decay, that much is expected from the Jordan-block form. The message is how it accelerates decay: not by retuning a pre-existing channel, but by collapsing the eigenbasis and removing the channel altogether. The slower decay channel does not simply become faster; it ceases to exist as an independent degree of freedom. This distinction, often obscured in steady-state descriptions, becomes sharply visible in the time domain. The square-root scaling of the decay rate provides the quantitative signature of this collapse. Its functional form is the same as that of the frequency splitting, a familiar hallmark of EPs, but measured here in the transient dynamics. This correspondence suggests that the Jordan-block geometry imprints itself on the time domain in a way that is as robust and universal as it is on the spectrum. The invariance of the fluorescence background reinforces this interpretation: the effect is not a local modification of emission rates, but a global restructuring of the coupled-state space, selective to coherent fields that establish long-range phase relations.

The in-situ reconfigurability of this effect, achieved by simply adjusting pump conditions without altering the device, distinguishes it from conventional approaches that require permanent structural changes. The extension to higher-order EPs in multi-cavity systems (see SI) points to a family of such topological dissipation effects, with scaling exponents that increase with the order of the EP. Beyond the specific system studied here, the ability to manipulate transient decay through state-space topology may offer a new route for applications where the temporal response of photonic devices is the critical figure of merit, such as ultrafast switching, optical logic, and reconfigurable photonic networks.

## Acknowledgement

The authors acknowledge support by National Key Research and Development Program of China (Grant Nos. 2024YFB2809200 and 2022YFA1404700), National Natural Science Foundation of China (Grant Nos. 12574413, 12334016, U25A6014 and 62305084), Guangdong Basic and Applied Basic Research Foundation (Grant Nos. 2023A1515011746, 2023B1212010003, and 2024B1515020060), Shenzhen Fundamental Research Projects (Grant No. GXWD 2022081714551), Guangdong Provincial Quantum Science Strategic Initiative (Grant Nos. GDZX2406002 and GDZX2406001), Shenzhen Science and Technology Program (Grant Nos. JCYJ20230807094401004 and JCYJ20241202123719025), the Fundamental Research Funds for the Central Universities (Grant Nos. HIT.OCEF.2024020, HIT.OCEF.2025002, and 2022FRFK01013), and Self-Planned Task (Grant No. SKLRS202602B) of State Key Laboratory of Robotics and Systems (HIT), Financial Support for Outstanding scientific and technological innovation Talents Training Fund in Shenzhen.

## Data availability

The data that support the findings in this study is available from the corresponding authors upon reasonable request.

## Competing interests

The authors declare no competing interests.

## Author contributions

C.H. and Q.S. conceived the idea and supervised the research. Y.C., X. J. and R.J. prepared the experimental materials. Y.C performed the simulation. Y.C, R.J. and K.W performed the experimental measurements. C. H. prepared the manuscript from all author’s contributions.

## Additional information

Details of Sample Preparation, Optical Setup, and theoratical analysis are includde in the Supplementary information.